\documentclass[a4paper,amsmath,superscriptaddress,twocolumn,showkeys,showpacs,prl,aps, 10pt]{revtex4-1}
\usepackage{color}
\usepackage{graphicx}

\newcommand{\dagga}{{\phantom{\dagger}}}

\newcommand{\be}{\begin{equation}}
\newcommand{\ee}{\end{equation}}
\newcommand{\bea}{\begin{eqnarray}}
\newcommand{\eea}{\end{eqnarray}}
\newcommand{\ba}{\begin{eqnarray*}}
\newcommand{\ea}{\end{eqnarray*}}
\newcommand{\up}{\uparrow}
\newcommand{\down}{\downarrow}
\newcommand{\eqn}[1]{(\ref{#1})}

\begin{document}

\title{ Kondo impurities in  nanotubes: the importance of being "in"}

\author{P. P. Baruselli}
\affiliation{SISSA, Via Bonomea 265, Trieste 34136, Italy}
\affiliation{CNR-IOM, Democritos Unit\'a di Trieste, Via Bonomea 265, Trieste 34136, Italy}

\author{A. Smogunov}
\affiliation{CNR-IOM, Democritos Unit\'a di Trieste, Via Bonomea 265, Trieste 34136, Italy}
\affiliation{ICTP, Strada Costiera 11, Trieste 34014, Italy}
\affiliation{Voronezh State University, University Square 1, Voronezh 394006, Russia}
\affiliation{present address: CEA Saclay, France}

\author{M. Fabrizio}
\affiliation{SISSA, Via Bonomea 265, Trieste 34136, Italy}
\affiliation{CNR-IOM, Democritos Unit\'a di Trieste, Via Bonomea 265, Trieste 34136, Italy}
\affiliation{ICTP, Strada Costiera 11, Trieste 34014, Italy}

\author{E. Tosatti}
\affiliation{SISSA, Via Bonomea 265, Trieste 34136, Italy}
\affiliation{CNR-IOM, Democritos Unit\'a di Trieste, Via Bonomea 265, Trieste 34136, Italy}
\affiliation{ICTP, Strada Costiera 11, Trieste 34014, Italy}

\date{\today}

\begin{abstract}
Transition metal impurities will yield zero bias anomalies in the conductance of well contacted 
metallic carbon nanotubes, but Kondo temperatures and geometry dependences have not been anticipated so far. 
Applying the density functional plus numerical renormalization group 
approach  of Lucignano \textit{et al.}  to Co and Fe impurities in (4,4) and (8,8) nanotubes, we discover a 
huge difference of behaviour between outside versus inside adsorption of the impurity. The predicted 
Kondo temperatures and zero bias  anomalies, tiny outside the nanotube, turn large and strongly radius 
dependent inside, owing to a change of symmetry of the magnetic orbital.  Observation of this
Kondo effect should open the way to a host of future experiments.
\end{abstract}

\maketitle


Nanotubes provide a rich playground for a variety of many body phenomena, in particular quantum 
transport between metal leads~\cite{mc_euen}. Depending on transparency of the electrical contact between 
the nanotube and the leads, conduction may range from insulating with strong Coulomb blockade for poor 
contacts~\cite{blockade}, to 
free 
ballistic transport with conductance close to	4$e^2 / h$ when contact 
transmission is close to one~\cite{ballistic, ballistic2}. 
Kondo effects in intrinsic nanotubes have been described, 
either for poor contacts~\cite{nanocontact_Kondo_old,nygard2000}, and/or in connection with superconducting leads~\cite{bouchiat}, but none of the classic, extrinsic, single-atom impurity type. 
Here we focus on a high transmission lead-nanotube-lead contacts, with a single magnetic impurity 
adsorbed inside or outside a metallic nanotube segment -- an extrinsic case. Conceptually, this should constitute a 
reproducible system, whose conductance can be precisely and predictably  controlled by standard external agents 
such as magnetic field, gate voltage and temperature.  The ballistic conductance of the four nanotube conduction channels
will be altered by Kondo impurity screening, showing up as a zero bias anomaly ~\cite{zero_bias_Kondo}, in a way and
to an extent which is presently unknown. Transport anomalies have long been reported~\cite{odom2000} 
in a tip-impurity-deposited nanotube geometry, at tip-impurity-metal systems~\cite{knorr2002, vitali2008}; and at 
molecular magnetic break junctions~\cite{Parks11062010} -- systems with very limited atomic and structural control. 
For an atomistically defined system like ours, we aim at theoretical predictions that are not just generic -- 
as is often the case in Kondo problems -- but quantitative and  precise about Kondo temperatures, conductance anomaly
widths and lineshapes.  For that purpose, we need to implement an {\sl ab-initio} based protocol.
According to the "DFT+NRG" formulation by Lucignano {\sl et al.}~\cite{lucignano2009} that goal can be achieved
by solving
a custom-built Anderson model whose parameters 
are determined by the first principles derived impurity scattering phase shifts. As a specific application, 
we examine here Co and Fe impurities adsorbed on the outside surface of metallic armchair SWNTs. Results are at first
disappointing, predicting exceedingly small Kondo temperatures, and tiny conductance anomalies 
that would be hard to observe. When adsorbed {\it inside} the nanotube however, the same impurities should  
yield order of magnitude larger Kondo temperatures, which moreover increase with decreasing nanotube radius. 
When inside, in fact, the impurity magnetic orbital symmetry switches from parallel to perpendicular to the tube axis, 
causing a dramatic increase of hybridization with the carbon $\pi$-orbitals, and a corresponding surge of Kondo energy. 

\begin{figure}[!htb]
\includegraphics[width=0.4\textwidth]{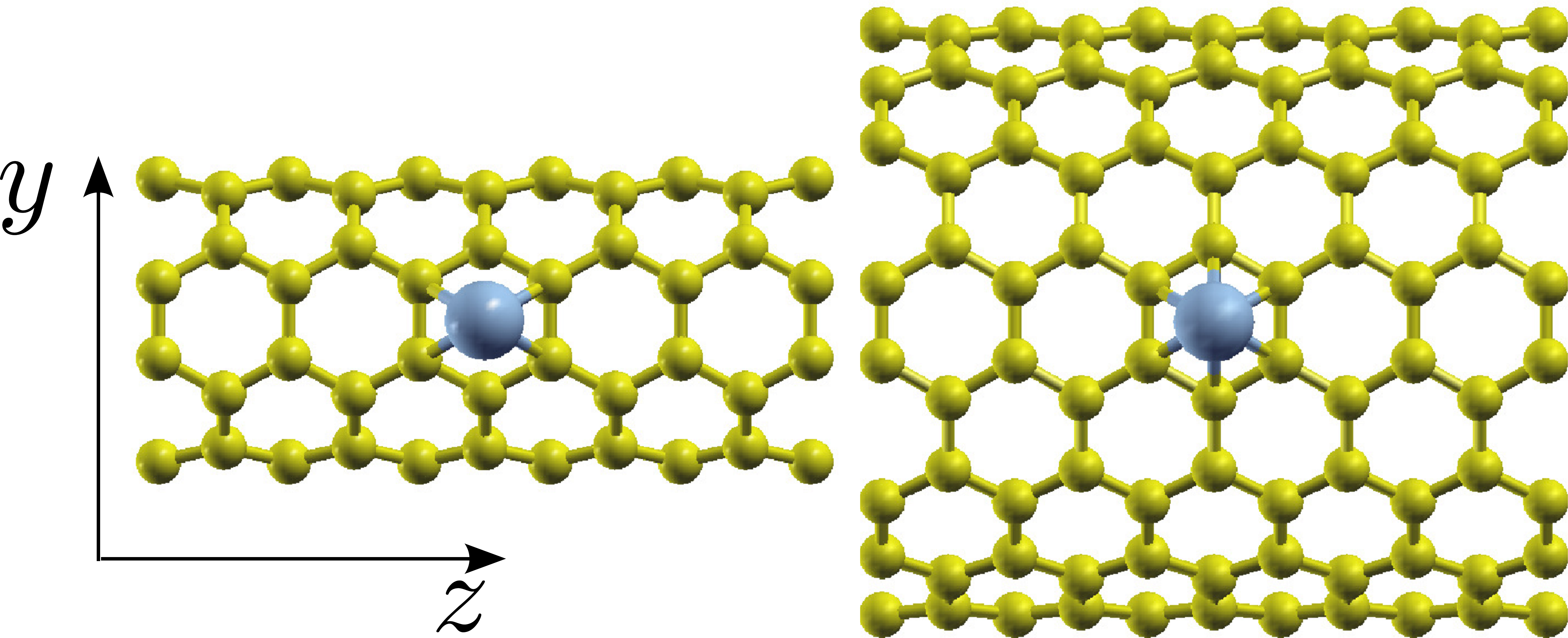}
\caption{Sketch of the (4,4) (left) and (8,8) (right) SWNTs in the $yz$ plane, with an impurity adsorbed in the hollow position (either inside or outside).}
\label{fig_struc}
\end{figure}
\begin{figure}[!htb]
\includegraphics[width=0.4\textwidth]{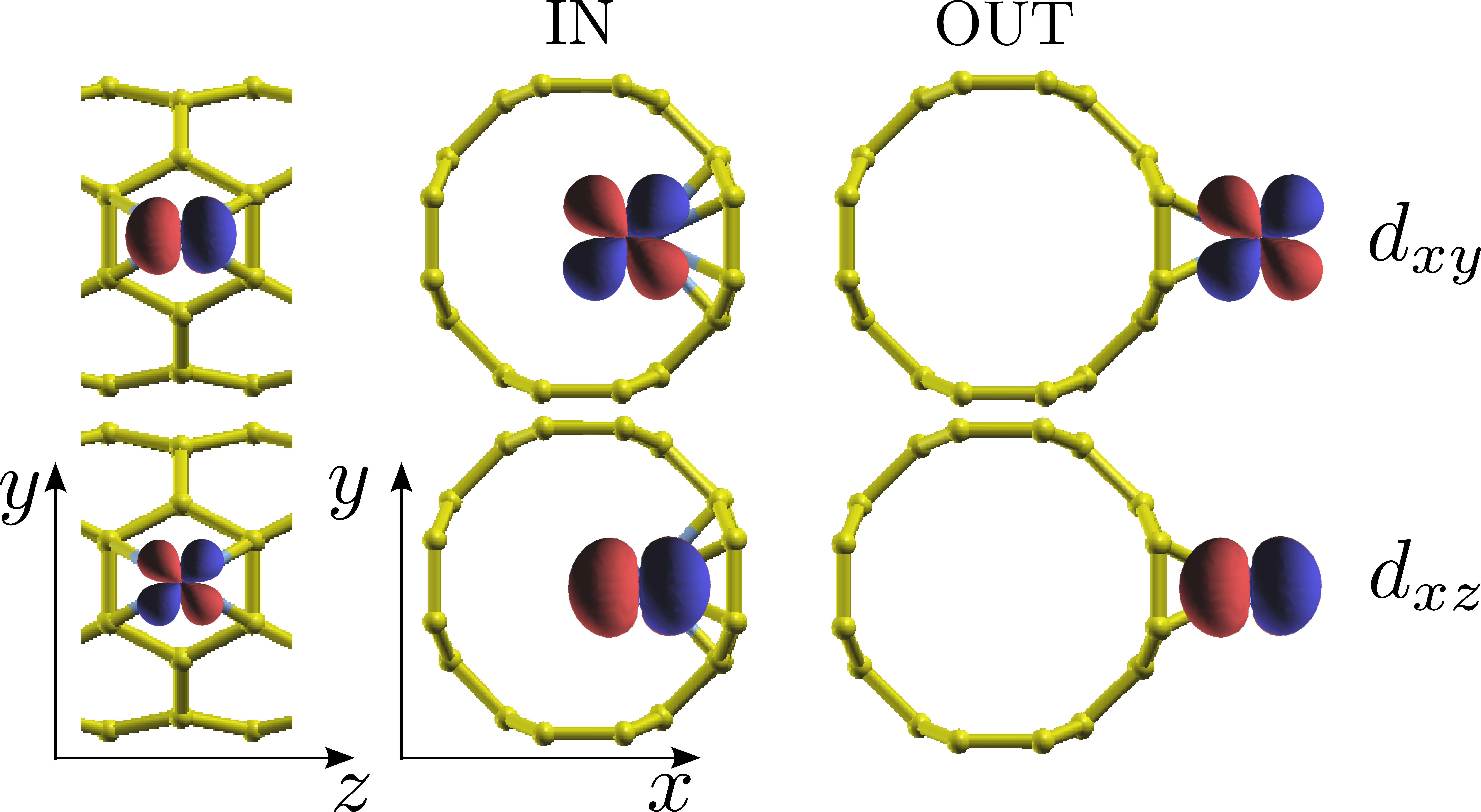}
\caption{Spatial distribution of $d_{xy}$ and $d_{xz}$ orbitals on the (4,4) SWNT (different colours mean a change of sign of the wavefunction).For Co, $d_{xy}$ (well hybridized) is the relevant orbital when inside, and $d_{xz}$ (poorly hybridized) when outside; for Fe, all orbitals are relevant.}
\label{fig_orb}
\end{figure}

Following Lucignano {\sl et al.}~\cite{lucignano2009} we first carry out a standard spin-polarized density 
functional theory (DFT) electronic structure calculation of the nanotube with one impurity; the conduction 
$\pi$-electron phase shifts extracted from that calculation are used to fix parameters of an Anderson model; 
the model is solved by Numerical Renormalization Group (NRG) to obtain Kondo temperatures;
finally, a  non equilibrium Green function technique (NEGF) yields the conductance near zero bias.  
We choose (4,4) and (8,8) single wall nanotubes (SWNT) (Figs.~\ref{fig_struc},~\ref{fig_orb}) with a Co or a Fe atom 
adsorbed at the hexagon center in a fully relaxed position alternatively outside or inside the tube (details 
provided in Supplemental Material).


\begin{figure}[!htb]
\includegraphics[width=0.4\textwidth]{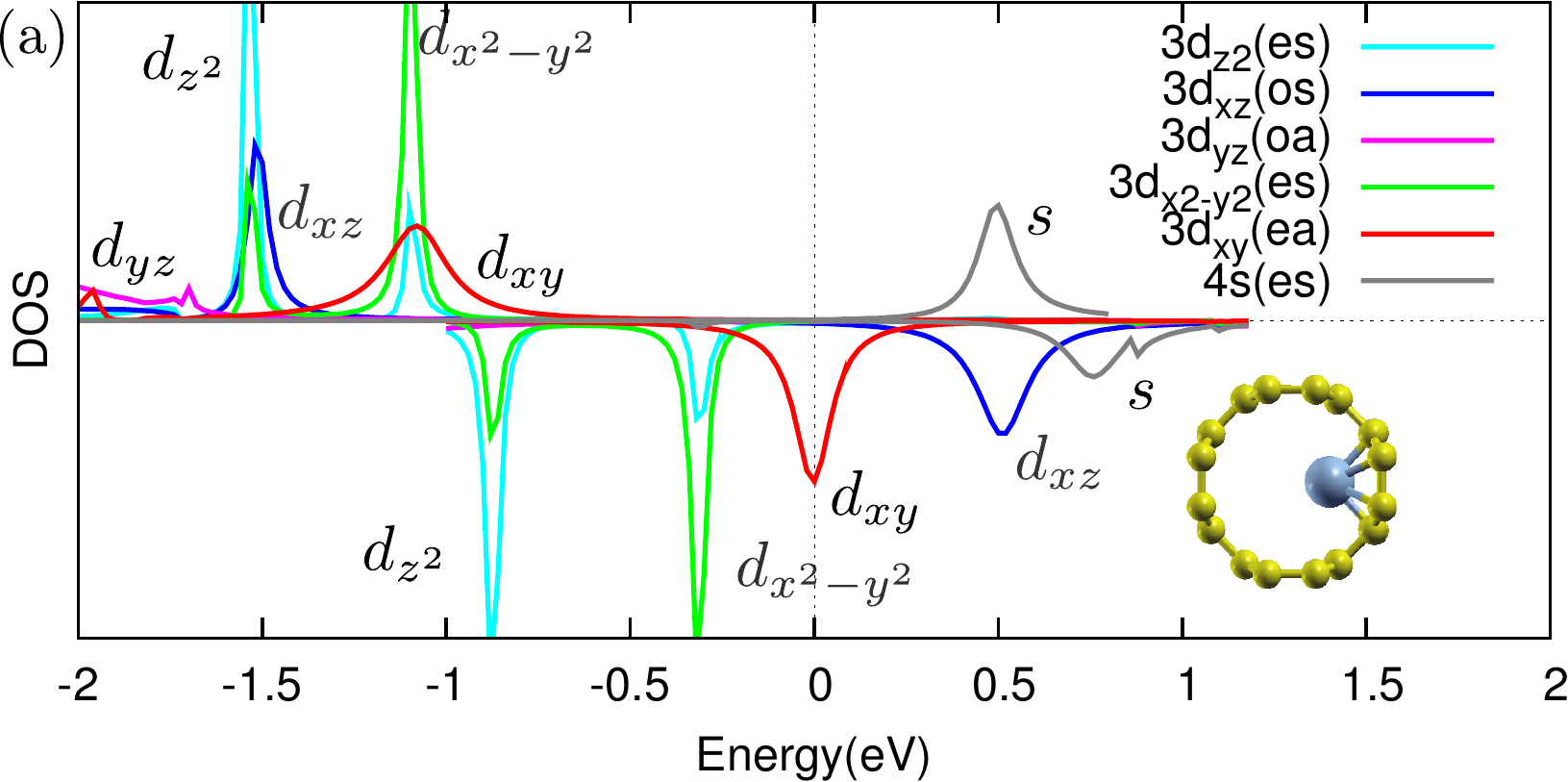}
\includegraphics[width=0.4\textwidth]{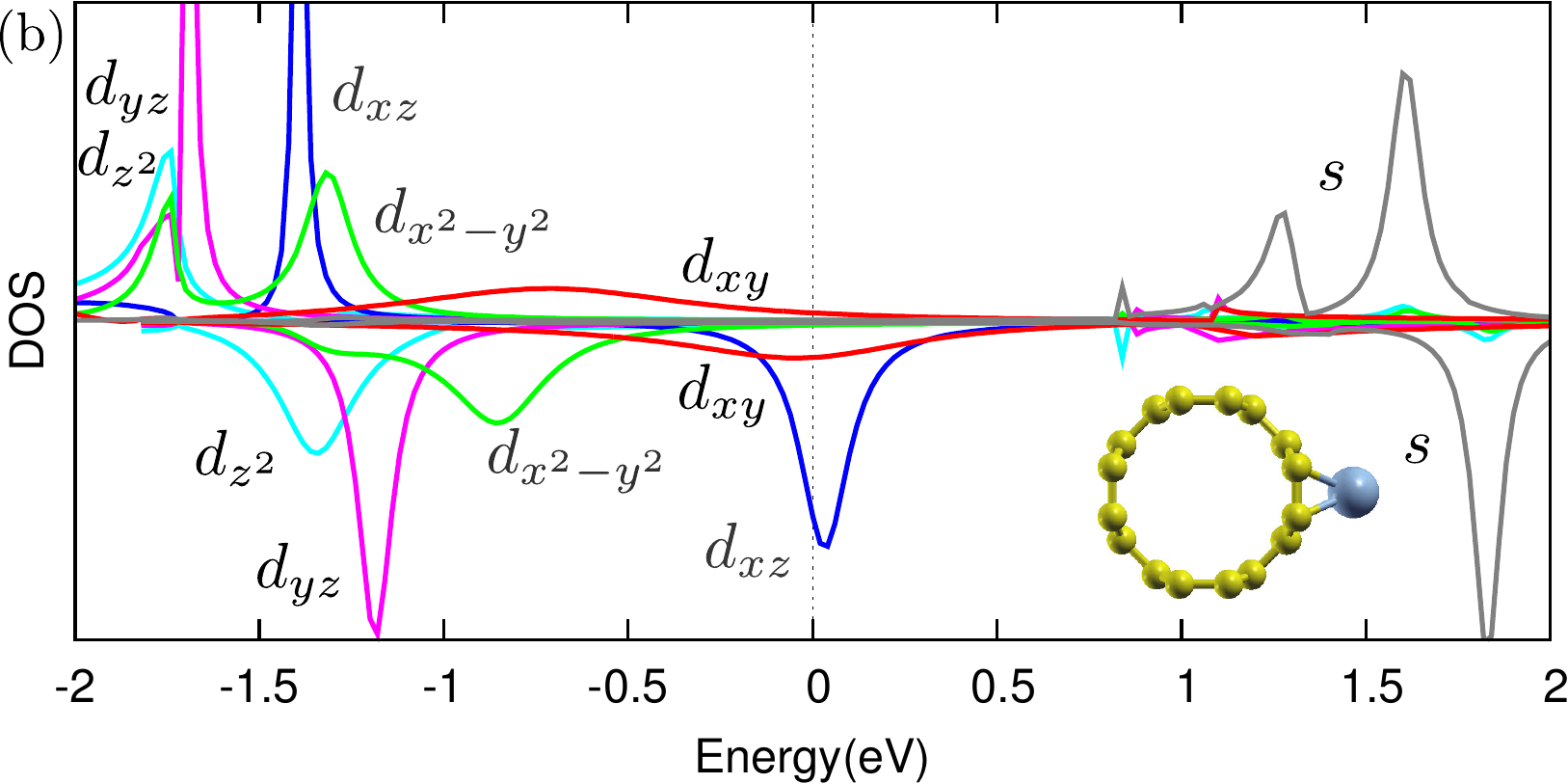}
\caption{
Symmetry resolved PDOS on the impurity atom for Co outside (a) and inside (b) the (4,4) SWNT. 
In the first case, orbital $d_{xz}$ is magnetic, while orbital $d_{xy}$ is weakly copolarized, and goes in the 
mixed-valence regime when the AIM is solved. In the second case, both orbitals are partly polarized; when 
the AIM is solved, orbital $d_{xy}$ goes to the Fermi energy, while orbital $d_{xz}$ is in the mixed-valence 
regime. Results are qualitatively the same on the (8,8) SWNT, where however energies differences between 
$d_{xz}$ and $d_{xy}$ orbitals are even smaller. When Fe is considered instead of Co, both orbitals are magnetic in all cases.}
\label{fig_pdos}
\end{figure}

The impurity projected density of electronic states is shown in Fig.~\ref{fig_pdos}.





DOS peaks mark the impurity $d$-states. Relative to the impurity site, states are even ($e$) or odd ($o$) 
under reflection across an $xy$ plane (orthogonal to the tube), and symmetric ($s$) or asymmetric ($a$) under
reflection across an $xz$ plane slicing the tube lengthwise.  Note the "magnetic" orbitals, where up and down
spins are exchange-split below and above the Fermi level respectively. In Co/(4,4)--OUT there is a single magnetic 
orbital $d_{xz}$ with $\{o,s\}$ symmetry 
indicating a S=1/2 state for Co ($3d^9 4s^0$), and S=1 for Fe ($3d^84s^0$) on (8,8). 
Consider connecting the two ends of  a nanotube segment to metal leads, and passing a current. If the contacts
are transparent, ballistic transport along the metallic nanotube will take place through the two bands at Fermi
(see  Fig.~\ref{fig_bands},  Supplemental Material).  
Left- and right-moving electronic states, $\phi_l$ and $\phi_r$ give rise in $e$ and $o$ combinations, 
$\phi_{e/o}=(\phi_l\pm\phi_r)/\sqrt2$ to four channels with distinct symmetries $\{e/o,s/a\}$ implying
without impurities a conductance $4e^2/h$ for perfectly transmitting contacts. 
A single impurity will cause each conduction channel to scatter onto the impurity orbital(s) of same symmetry, if any,
giving rise to a scattering phase shift. 
The  $(8\times8)$ unitary $S$-matrix 
is diagonal with eigenvalues  $e^{2i\delta_{\mu\sigma}}$ in the $\{e/o,s/a\}$ representation,  
where $\mu = es, ea, os, oa $, $\sigma=\uparrow$, $\downarrow$. 
The transmission and reflection probabilities
$|t_{\alpha\sigma}|^2 = \cos^2(\delta_{e\alpha\sigma}-\delta_{o\alpha\sigma}),
~|r_{\alpha\sigma}|^2 = \sin^2(\delta_{e\alpha\sigma}-\delta_{o\alpha\sigma})$, $\alpha=s,a$, also relate via the 
Friedel sum rule $\Delta \rho_{\alpha\sigma}(E) = \frac{1}{\pi} \frac{d \delta_{\alpha\sigma}(E)}{dE}$ 
to the extra DOS $\Delta \rho_{\alpha\sigma}$ induced by the impurity for each symmetry and spin. 
Phase shifts calculated by DFT are then used to determine
the parameters of an Anderson impurity model for the impurity. For each channel, 
we introduce spin rotation angles defined as $\theta_{\mu}=2(\delta_{\mu\downarrow}-\delta_{\mu\uparrow})$.
The most general AIM should include four scattering channels, $i=es,ea,os,oa$,  
and six impurity orbitals (one $s$ and five $d$), $a=1,\dots,6$, hence it is of the form
\bea
H &=& \sum_{ik\sigma}\,\Bigg(\epsilon_{k}\,c^\dagger_{ik\sigma}c^\dagga_{ik\sigma} + 
\sum_a\,V_{ik,a}\,\left(c^\dagger_{ik\sigma}d^\dagga_{a\sigma} + H.c.\right)\Bigg)\nonumber \\
&& + \sum_{ikk'\sigma}\,t_{i,kk'}\,c^\dagger_{ik\sigma}c^\dagga_{ik'\sigma} + H_{imp},\label{Ham-full}
\eea
where $c^\dagger_{ik\sigma}$ creates a spin $\sigma$ electron in channel $i$ with momentum $k$ along the tube, 
$d^\dagger_{a\sigma}$ a spin $\sigma$ electron in the orbital $a$ of the impurity. 
$V_{ik,a}$ is the hybridization matrix element between conduction and impurity orbitals, which is finite 
only if they share the same symmetry
, while $t_{i,kk'}$ describes 
a local scalar potential felt by the conduction electrons because of the translational symmetry breaking 
caused by the impurity. $H_{imp}$ includes all terms that involve only the impurity orbitals, which, 
since the orbital $O(3)$ symmetry is fully removed by crystal field, can be written as  
\bea
H_{imp} &=&  \sum_{a\sigma}\,(\epsilon_a\,n_a + U_a\,n_{a\up}\,n_{a\down})\nonumber \\
&& +\sum_{a<b}\,U_{ab}\,n_a\,n_b + 2J_{ab}\,\mathbf{S}_a\cdot\mathbf{S}_b,\label{H-imp}
\eea
where $n_{a\sigma}=d^\dagger_{a\sigma}d^\dagga_{a\sigma}$, $n_a=\sum_\sigma\,n_{a\sigma}$ and 
$J_{ab}<0$, favoring a ferromagnetic correlation among the spin densities $\mathbf{S}_a$ of 
the different orbitals. 
The parameters of this Hamiltonian 
are fixed by 
requiring them to reproduce in the mean field approximation the ab initio DFT shifts, in addition to orbital energies~\cite{lucignano2009,baruselli2011}.
The AIM hamiltonian \eqn{Ham-full} found in this way is still numerically prohibitive. 
Since our ultimate goal is transport at low temperature and small bias, we can  neglect 
orbitals that within DFT are either doubly occupied or empty. 
In this approximation there are two active magnetic orbitals for both Co and for Fe, either outside or inside the tube. 

The crucial difference between outside and inside arises in the nature
of the magnetic orbital. For Co outside, $d_{xz}$ has $os$ symmetry, lying in the $y=0$  plane (see Fig. \ref{fig_orb}). 
Its hybridization, 
\[
\Gamma_{os,xz} = \pi\sum_k\,V_{osk,xz}^2\,\delta\left(\epsilon_k-\epsilon_F\right),
\] 
the controlling parameter of the Kondo effect, is small. When Co is inside on the contrary, the magnetic orbital
is $d_{xy}$, lying in the $z=0$ plane, its radial lobes (see Fig. \ref{fig_orb}) much more hybridized with the tube conduction channels.
The orbital switching between outside and inside is due to a reversal of crystal field, and to a different hybridization. 
In the case of Fe instead 
both $d_{xz}$ and $d_{xy}$ are magnetic, and the change in crystal field does not play any major role. 
In both Co and Fe, the tangential nature of the outside magnetic orbital implies no strong dependence of hybridization
upon tube radius. Conversely, the radial nature of the magnetic orbital gives rise to a large radius dependence when the imprity is inside,
where it is better "surrounded"  especially for smaller tube radius. 
The impurity-nearest carbon coupling $V$ leads to a hybridization width 
for a $(n,n)$ nanotube,  
$\Gamma\sim \pi V^2\rho/n \propto 1/n$  where $\rho$ is the (radius independent) density of states at the Fermi energy. 
This increase of $V$ with inverse radius explains the increased coupling of orbital $d_{xz}$, for example, of Co on tube  $(4,4)$ (0.087 eV) with 
respect to$(8,8)$ (0.058 eV), only partly compensated by  a 
slight 
decrease of $V$ due to the larger curvature.
As a consequence, Kondo temperatures are predicted to decrease exponentially with increasing tube radius -- so long as higher 
subbands can be neglected. For very large tubes, our single subband model is invalid and higher 
subbands must be taken into account.

It is worth here discussing, at least qualitatively, the limit of zero curvature,  graphene.
For Co or Fe on graphene the orbitals $d_{xz}$ and $d_{xy}$ are degenerate, and occupied by three electrons only. 
This unstable SU(4) symmetry can be broken by e.g., spin-orbit~\cite{wehling}, or by a Jahn-Teller distortion, 
both leading to an ordinary SU(2) Kondo effect. 
In either case, Co/nanotube Kondo is basically different from Co/graphene.~\cite{wehling}
For Fe/nanotube, with two electron in two orbitals, the additional Hund's rule coupling, of order 1 eV , is 
larger than the crystal-field splitting, hybridization differences, and spin-orbit interaction -- hence both orbitals should 
jointly undergo Kondo screening. The same conclusion should apply to Fe/graphene even though the two 
orbitals become degenerate, because the Hund's exchange forces the two electrons to occupy each a different orbital 
in a spin-triplet configuration.

\begin{figure}[!htb]
\includegraphics[width=0.4\textwidth]{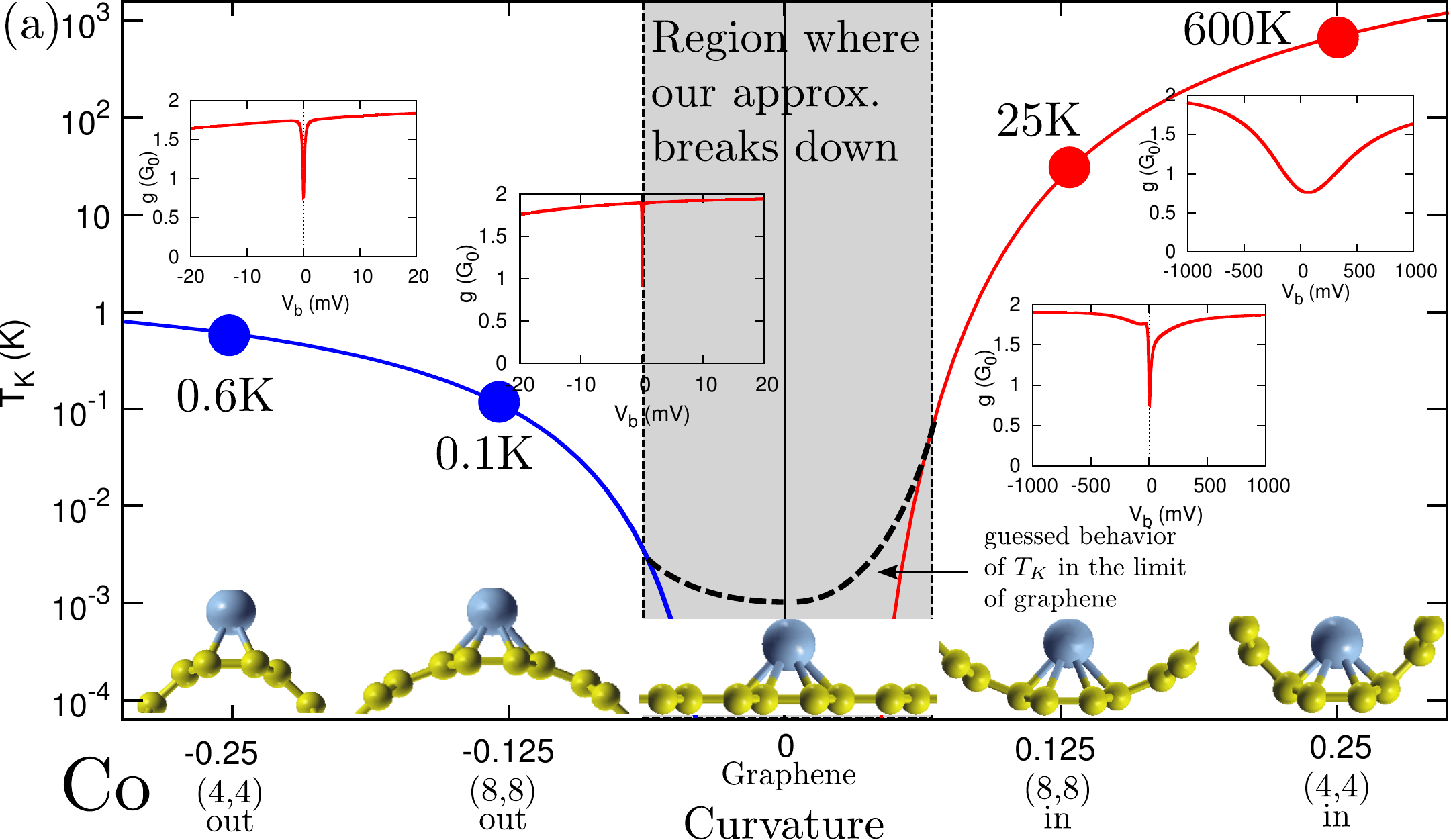}\\
\vspace{0.5cm}
\includegraphics[width=0.4\textwidth]{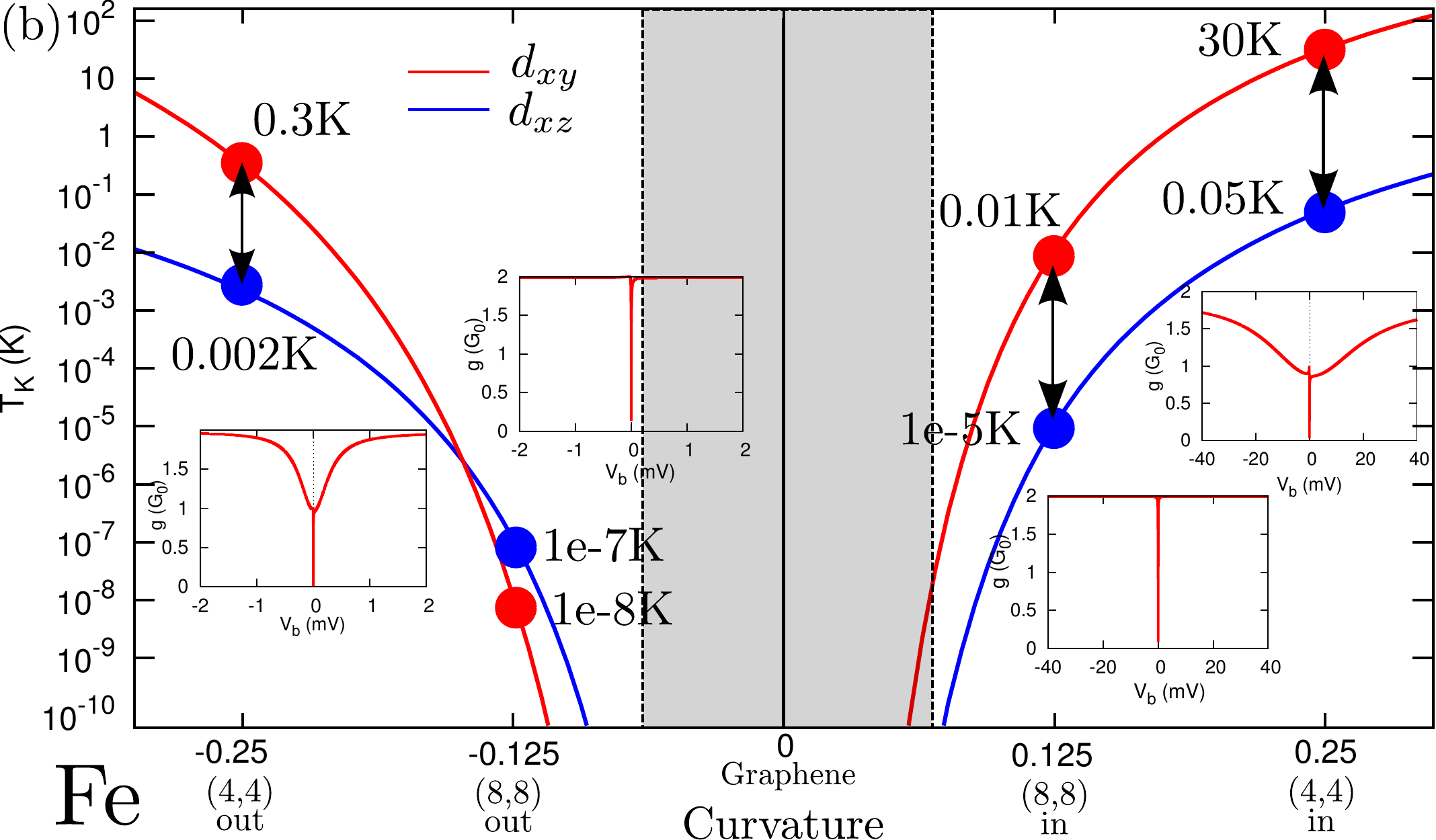}
\caption{Kondo temperatures as a function of curvature for Co (a) and Fe (b); predixted zero bias anomalies are also shown for each case.  Dots show calculated values from tab. \ref{table1}, lines are best fits assuming (independently for negative and positive curvatures) $log(T_K)=a-b/|x|$ where $x$=curvature. The shaded area shows the region of low curvature, where our single band approximation breaks down, and additional terms, such as spin-orbit coupling, must be included in the Hamiltonian. As a consequence of considering higher subbands, the Kondo temperature, that in our model goes exponentially to zero, can saturate at a finite value (dotted line in fig. (a), not shown in fig. (b); the saturation value is guessed). In fig. (b), an arrow shows the range of temperatures where Fe behaves like an underscreened impurity.}
\label{fig_tk}
\end{figure}


The simplified Anderson impurity models just obtained are solved by standard Numerical Renormalization 
Group (NRG)~\cite{bulla08}; we adopt a two-band model, which takes into account channels $os$ and $ea$, and 
orbitals $d_{xy}$ and $d_{xz}$. The approximate Kondo temperatures obtained in this manner (details
in Supplemental Material) are given in Table \ref{table1}, column 6. We should warn here that Kondo temperature 
are by construction affected by a large error, because of their intrinsic exponential dependence on parameters. With that caveat,
we 
verify, as already stated, that Kondo temperatures turn from very small when impurities are outside the nanotube, to large
and radius dependent when inside. The impurity inside the nanotube is therefore the geometry which we propose for experimental verification.

\begin{table}[tb]
 $$
\begin{array}{|c|c|c|c|c|c|c|}\hline
\mbox{Impurity}&\mbox{Nanotube}&\mbox{Position}&\mbox{Orbital}&\Gamma&T_K(K)&q\\\hline
\mbox{Co}	&(4,4)	&\mbox{Out}&d_{xz}&0.087&	0.6&	-0.03\\
\mbox{Co}	&(8,8)	&\mbox{Out}&d_{xz}&0.058&	0.1&	-0.04\\
\mbox{Co}	&(8,8)	&\mbox{In}
&d_{xy}&0.126&25	&-0.10	\\
\mbox{Co}	&(4,4)	&\mbox{In}
&d_{xy}&0.380&600	&	-0.11\\\hline
\mbox{Fe}	&(4,4)	&\mbox{Out}&d_{xz}&0.092&	0.002	&0.01\\
		&	&		&d_{xy}&0.082&0.3	&	-0.02\\
\mbox{Fe}	&(8,8)	&\mbox{Out}&d_{xz}&0.062&	10^{-7}	&0.06\\
		&	&		&d_{xy}&0.044&10^{-8}	&	-0.02\\
\mbox{Fe}	&(8,8)	&\mbox{In}&d_{xz}&0.081&10^{-5}		&0.09\\
		&	&		&d_{xy}&0.134&0.01	&	-0.01\\
\mbox{Fe}	&(4,4)	&\mbox{In}&d_{xz}&0.126&0.05	&0.35\\
		&	&		&d_{xy}&0.396&30	&-0.06	\\\hline
\end{array}
$$
\caption{Kondo orbitals for each system, with broadening $\Gamma$ and Kondo temperature $T_K$. }
\label{table1}
\end{table}


The zero-bias conductance anomalies and lineshapes 
are calculated by 
the Keldysh method for non-equilibrium Green functions~\cite{meir}. Approximating the Kondo 
resonance by a Lorentzian, the conductance for a single channel is a Fano resonance
$ g(v)=\frac{(q+v)^2}{(q^2+1)(v^2+1)}$ where $v\equiv V_{B}/\Gamma_K$ is the dimensionless bias potential, 
$q$ is the shape parameter, and total conductance is the sum of even and odd channels. 
Results are shown in Table~\ref{table1} and Fig.~\ref{fig_tk}. 
We generally predict in all cases a conductance minimum at zero bias ($q\simeq 0$), 
except in the case of Co inside the (4,4) tube, where hybridization is so large that the Kondo effect gives 
way to a frank resonant level.
Finally, the Fe impurity, having two different Kondo temperatures can, especially when inside, behave 
as an underscreened impurity in the range of temperatures for which $T_{K1} \ll T \ll T_{K2}$. 


In summary, we applied a DFT + NRG method to obtain first principle predictions of the Kondo effect 
in the conductance of well contacted metallic nanotube segments hosting a single transition metal impurity atom.
One first novelty is our claim of ab initio predicting power, which has not been common in Kondo problems. 
The main physical surprise is a strong difference between outside and inside impurity adsorption; only when inside, the 
Kondo temperatures are large and increase for decreasing radius. The large radius limit, graphene, is shown to
be an intrinsically different case.  Experimentally, it should be possible
to insert an impurity inside a metallic nanotube segment, long enough that the finite size level discreetness 
is smaller than Kondo energies, and short enough for strong correlations to be negligible.  The predicted Kondo 
temperatures and zero bias anomalies for small radius nanotubes are very substantial and should not only
be measurable, but should open the way to a variety of new observations in a larger variety of metallic nanotubes.

\begin{acknowledgments}
This work was supported by PRIN/COFIN  20087NX9Y7.
P. P. Baruselli would like to thank L. De Leo for providing the NRG code and for useful discussions. 
\end{acknowledgments}

\bibliographystyle{apsrev}

\end{document}